\documentclass[a4paper,11pt]{article}
\pdfoutput=1 

\usepackage{jcappub} 

\usepackage[T1]{fontenc} 
\usepackage{footnote}
\usepackage[bottom]{footmisc}
\usepackage{graphicx}
\usepackage{epsfig}

\usepackage{mwe}
\usepackage{subfig}

\newcommand{\beq}{\begin{equation}}
\newcommand{\eq}{\end{equation}}
\newcommand{\bear}{\begin{eqnarray}}
\newcommand{\ear}{\end{eqnarray}}

\newcommand{\be}{\begin{equation}}
\newcommand{\ee}{\end{equation}}
\newcommand{\bea}{\begin{eqnarray}}
\newcommand{\eea}{\end{eqnarray}}

\begin{document}


\title{E-ELT constraints on runaway dilaton scenarios}

\author[a]{M. Martinelli,} 
\emailAdd{m.martinelli@thphys.uni-heidelberg.de}
\author[b]{E. Calabrese,}
\emailAdd{erminia.calabrese@physics.ox.ac.uk}
\author[c,d]{C.J.A.P. Martins}
\emailAdd{carlos.martins@astro.up.pt}

\affiliation[a]{Institut f\"ur Theoretische Physik, Ruprecht-Karls-Universit\"at Heidelberg, Philosophenweg 16, 69120, Heidelberg, Germany}
\affiliation[b]{Sub-department of Astrophysics, University of Oxford, Keble Road, Oxford OX1 3RH, UK}
\affiliation[c]{Centro de Astrof\`isica, Universidade do Porto, Rua das Estrelas, 4150-762 Porto, Portugal}
\affiliation[d]{Instituto de Astrof\`isica e Ciencias do Espaco, CAUP, Rua das Estrelas, 4150-762 Porto, Portugal}

\abstract{
We use a combination of simulated cosmological probes and astrophysical tests of the stability of the fine-structure constant
$\alpha$, as expected from the forthcoming European Extremely Large Telescope (E-ELT), to constrain the class of string-inspired runaway dilaton models of Damour, Piazza and Veneziano. 
We consider three different scenarios for the dark sector couplings in the model and discuss the observational differences between
them. We improve previously existing analyses investigating in detail the degeneracies between the parameters ruling the 
coupling of the dilaton field to the other components of the universe, and studying how the constraints on these parameters change for different fiducial cosmologies. We find that if the 
couplings are small (e.g., $\alpha_b=\alpha_V\sim0$) these degeneracies strongly affect the constraining power of future data, while if they are sufficiently large 
(e.g., $\alpha_b\gtrsim10^{-5}-\alpha_V\gtrsim0.05$, as in agreement with current constraints) the degeneracies can be partially broken. We show that E-ELT will be able to probe some of 
this additional parameter space.}

\date{\today}
\maketitle

\section{Introduction}
\label{sec:intro}

Cosmology and particle physics are in the exciting period in which their standard theoretical models can be precisely tested with a plethora of high-resolution data. 
However there are strong hints that neither model is complete, and that possible new physics may be within the reach of the next generation of probes.

Despite the success of the standard model of particle physics, highlighted by the confirmation of the long-sought Higgs
particle, there are three firmly established facts that it can't yet explain: neutrino masses, dark matter and the size of
the baryon asymmetry of the universe. Crucially, all three have cosmological implications.

Cosmological observations of the universe's large-scale-structure and microwave relic radiation -- the Cosmic Microwave Background, CMB -- strongly support the predictions of the 
so-called Hot Big-Bang model and a late-time accelerated cosmic expansion \cite{Ade:2015xua}. The current cosmological paradigm (known as $\Lambda$CDM) is based on three pillars: inflation, 
dark matter and dark energy, each of which relies on currently unknown physics. 

In particular, the precise nature of dark energy is one of the deepest enigmas of modern physics. The standard cosmological model recognizes dark energy as the dominant component of the energy 
budget of the universe today and identifies it as the one responsible for the cosmic acceleration. The next generation of astrophysical facilities must strive to search for, identify and 
ultimately characterise this mysterious component. The most obvious task to start identifying 
dark energy is to ascertain whether this is due to a cosmological constant (as introduced by Einstein) or to a new dynamical degree of freedom. 

It is common to model a dynamical dark energy through a scalar field (see e.g., \cite{Kunz2012}); a connected task is therefore to identify theories 
which lead to the addition of new degrees of freedom and thus can produce this scalar field. An interesting example is the scalar partner of the spin-2 graviton,
known as the dilaton, hereafter denoted $\phi$, whose existence is predicted by string theory.\\

Here we study in 
detail the cosmological consequences of a particular class of string-inspired models, the runaway dilaton scenario of Damour, Piazza and Veneziano \cite{Damour:2002mi,Damour:2002nv}.
The conceptual appeal of this scenario stems from the fact that it may provide a way to reconcile a massless dilaton with experimental data. On the other hand, from the observational 
point of view its distinguishing feature is the fact that, with significant dark sector couplings, these models yield violations of the Equivalence Principle and variations of the 
fine-structure constant $\alpha$ that are potentially measurable. With the recent progress in experimental tests of the former (summarized in \cite{Wagner:2012ui}) and astrophysical 
tests of the latter (summarized in \cite{Ferreira:2015xea}), as well as the availability of high-resolution cosmological datasets, it is now possible to quantitatively test these models.

This work will complement the analysis presented in \cite{Martins:2015dqa} by investigating the gain in sensitivity (with respect to current data constraints) provided by forthcoming 
facilities, focusing on the European Extremely Large Telescope (E-ELT): its high-resolution ultra-stable spectrograph (known as ELT-HIRES) will significantly improve tests of the stability 
of fundamental couplings and will also be sensitive enough to carry out a first measurements of the redshift drift signal deep in the matter dominated era \cite{HIREStlr}. We explore the 
degeneracies that might arise between both cosmological and different dilaton model coupling parameters by letting them all free to vary --we note that this was not the case in 
\cite{Martins:2015dqa} where cosmological parameters were held fixed. We want to stress that exploring the whole parameters space is essential: we will show how degeneracies between 
parameters can be a strong limitation for this kind of models.

The paper is organized as follows. In Section \ref{sec:theo} we present the essential features of the runaway dilaton model
and study their impact on cosmological observables provided by the E-ELT. The method used to simulate the observables is described in Section \ref{sec:ana}; the analysis and the results 
are presented in Section \ref{sec:res}. Finally, Section \ref{sec:conc} reports the main conclusions and a discussion of some future prospects.

\section{Runaway dilaton cosmology}
\label{sec:theo}

The runaway dilaton scenario was first introduced in \cite{Damour:2002mi} and further developed in \cite{Damour:2002nv}. The latter also includes some simple cosmological constraints, 
which we updated in \cite{Martins:2015dqa} and will extend further here. In this Section we will introduce the features of the model that will be relevant for our analysis; we 
refer the reader to the original works for additional details.

The main motivation for this scenario is to reconcile a massless dilaton with experimental data (which we will further discuss shortly). This is achieved by having the dilaton decouple 
while cosmologically attracted towards an infinite bare coupling, with the coupling functions having a smooth finite limit:
\begin{equation}
B_i(\phi) = c_i + {\cal O}(e^{-\phi})\,,
\end{equation}
where the $c_i$ are constants. 
The Einstein frame Lagrangian is \cite{Damour:2002mi,Damour:2002nv}:
\begin{equation}
\mathcal{L}=\frac{R}{16\pi G}-\frac{(\nabla\phi)^2}{8\pi G}-\frac{1}{4}B_F(\phi)F^2+...\ ,
\end{equation}
where $R$ is the Ricci scalar, $F$ is the electromagnetic tensor and $B_F$ is the gauge kinetic function (which will determine the evolution of $\alpha$).\\
From the above one can derive the modified first Friedmann equation:
\begin{equation}\label{eq:fried1}
3H^2=8\pi G\sum_i{\rho_i}+H^2\phi'^2 \,,
\end{equation}
the sum here is extended over all the standard components of the universe and includes also the potential part of the scalar field, while the contribution of the kinetic term is 
considered in the last term (in which ${}'$ denotes the derivative with respect to the logarithm of the scale factor $a$).

The total energy density and pressure of the field are in effect expressed as a sum over the kinetic ($k$) and potential (V) parts of the field: \\
\begin{equation}
\rho_\phi=\rho_k+\rho_V=\frac{H^2\phi'^2}{8\pi G}+V(\phi)
\end{equation}
\begin{equation}
p_\phi=p_k+p_V=\frac{H^2\phi'^2}{8\pi G}-V(\phi).
\end{equation}
Therefore the dilaton contributes to the cosmological expansion as a an effective quintessence-like field through its
potential (from now on we will assume this contribution to be equivalent to a cosmological constant) with a correction
brought to the Friedmann equation by its kinetic part. However, as we will now see, this class differs from simple
canonical quintessence models, since the dilaton does couple to the rest of the model.

From the same Lagrangian, we can obtain the equation of motion for the field $\phi$, which reads:
\begin{equation}\label{eq:field}
\frac{2}{3-\phi'^2}\phi''+\left(1-\frac{p}{\rho}\right)\phi'=-\sum_i{\alpha_i(\phi)\frac{\rho_i-3p_i}{\rho}}
\end{equation}
where $\rho$ and $p$ are, respectively, the total density and pressure obtained summing over the 
standard components of the universe and the potential part of the field. The $\alpha_i$ are free functions 
which characterize the coupling of the field with the different components, i.e., baryons, dark matter,
and effective dark energy. In what follows we will assume the coupling to the effective
dark energy $\alpha_V$ to be constant, while the coupling to baryonic matter is \cite{Damour:2002nv}:
\begin{equation}
\frac{\alpha_b(\phi)}{\alpha_{b,0}}=e^{-(\phi(z)-\phi_0)}\,,
\end{equation}
where the subscript $0$ indicates quantities evaluated at present time. We note that in the previous work \cite{Martins:2015dqa} this was denoted $\alpha_{\rm had}$.\\
As in \cite{Martins:2015dqa} we will consider three different possibilities for the dark matter coupling $\alpha_m$:
\begin{itemize}
\item "Dark Coupling" : $\alpha_m=\alpha_V$
\item "Matter Coupling" : $\alpha_m(\phi)=\alpha_b(\phi)$
\item "Field Coupling" : $\alpha_m(\phi)=-\phi'$ 
\end{itemize}
These three phenomenological choices, all motivated by the discussion in \cite{Damour:2002nv}, are meant to represent the range of possible behaviours in 
this class of models. The dark coupling corresponds to the simplest assumption, namely that the dark sector is characterized by as single coupling, appliccable to both dark 
matter and dark energy. Conversely the matter coupling assumes that there's a single coupling for dark matter and baryons while dark energy couples differently. Finally the field 
coupling corresponds to the approximate matter-era solution discussed in  \cite{Damour:2002nv}. In all these cases, the dark matter coupling depends on either $\alpha_{b,0}$ or $\alpha_V$, 
which are therefore the only free parameters of the model considered here. 

As discussed in \cite{Martins:2015dqa}, local measurements
of gravitational light deflection \cite{Bertotti:2003rm} and tests of violations of the weak equivalence
principle \cite{Wagner:2012ui,Muller:2012sea} provide constraints on the coupling of the field to baryonic matter:
\begin{equation}
|\alpha_{b,0}|\leq10^{-4}\,
\end{equation}
while the constraints on couplings to the dark matter sector are much weaker \cite{Wagner:2012ui,Stubbs:1993xk}:
\begin{equation}
|\alpha_{m,0}|\leq1\,.
\end{equation}

However, from Eq.(\ref{eq:fried1}) and assuming a spatially flat universe it is also possible to relate the field
derivative to the deceleration parameter $q$ as:
\begin{equation}
\phi^{'2}_0=1+q_0-\frac{3}{2}\Omega_m.
\end{equation}
Thus, using the currently available limits for $q_0$ and $\Omega_m$ \cite{Neben:2012wc,Ade:2015xua}
it is possible to obtain a constraint on the current velocity of the field:
\begin{equation}\label{eq:vellim}
|\phi'_0|\leq0.3.
\end{equation}
This bound implies that we can approximate the field to be moving slowly at present time and 
therefore we can rewrite Eq.(\ref{eq:field}) as:
\begin{equation}\label{eq:incond}
\phi'_0=-\frac{\alpha_b\Omega_b+\alpha_m\Omega_c+4\alpha_V\Omega_V}{\Omega_b+\Omega_c+2\Omega_V} \,,
\end{equation}
where $\Omega_b$ and $\Omega_c$ are the baryon and cold dark matter densities, and $\Omega_V$ is the effective dark energy density generated by the potential part of the field.

As shown in \cite{Martins:2015dqa}, combining this relation with the expression of $\alpha_m$ in terms of the other two couplings we can obtain conservative constraints on the dark energy 
coupling:
\begin{equation}
 |\alpha_V|\lesssim0.15.
\end{equation}

Eq. (\ref{eq:incond}) also provides an initial condition for Eq.(\ref{eq:field}). Moreover, the cosmological observables that we will consider (as shown in the next Section) depend only on 
either the field derivative $\phi'$ or the field shift $\phi(z)-\phi_0$, and we can solve Eq.(\ref{eq:field}) for $\phi(z)-\phi_0$ withouth loss of generality, setting the initial condition 
on the shift to zero.

\begin{figure*}[t]
\begin{center}
\includegraphics[width=7.65cm]{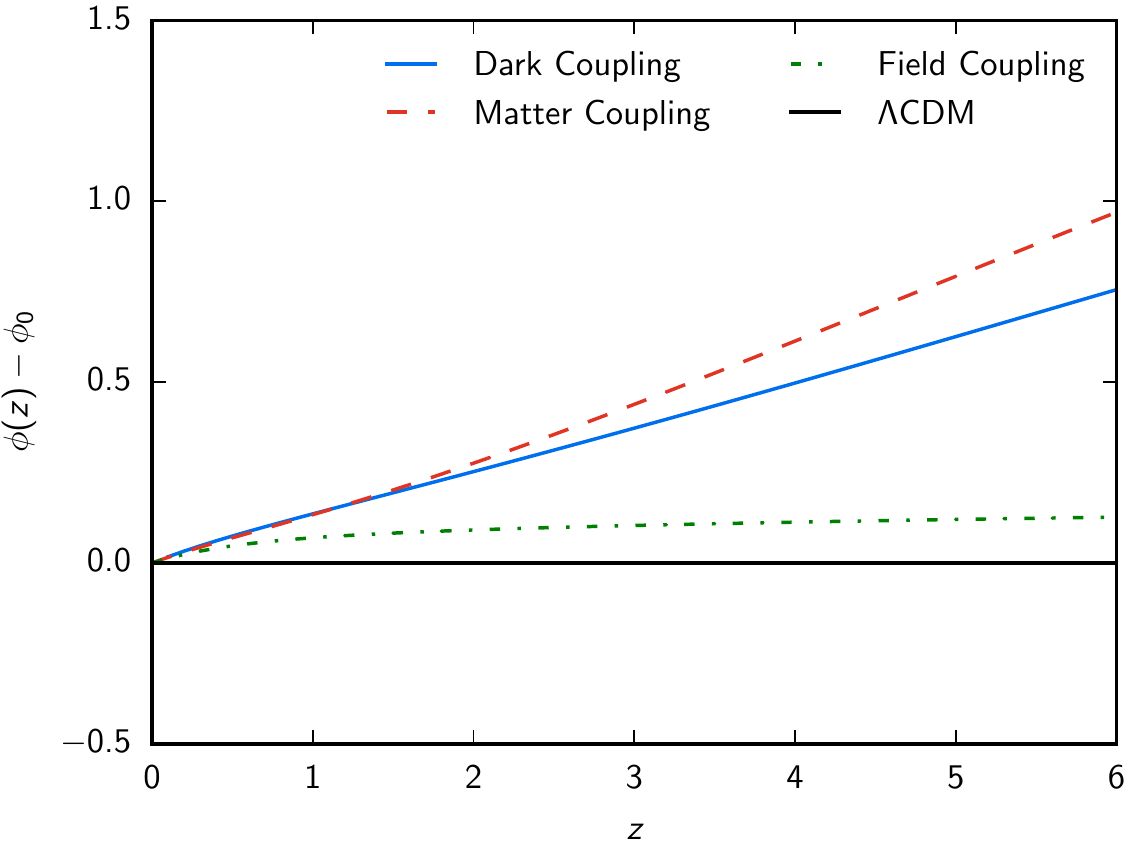} 
\includegraphics[width=7.65cm]{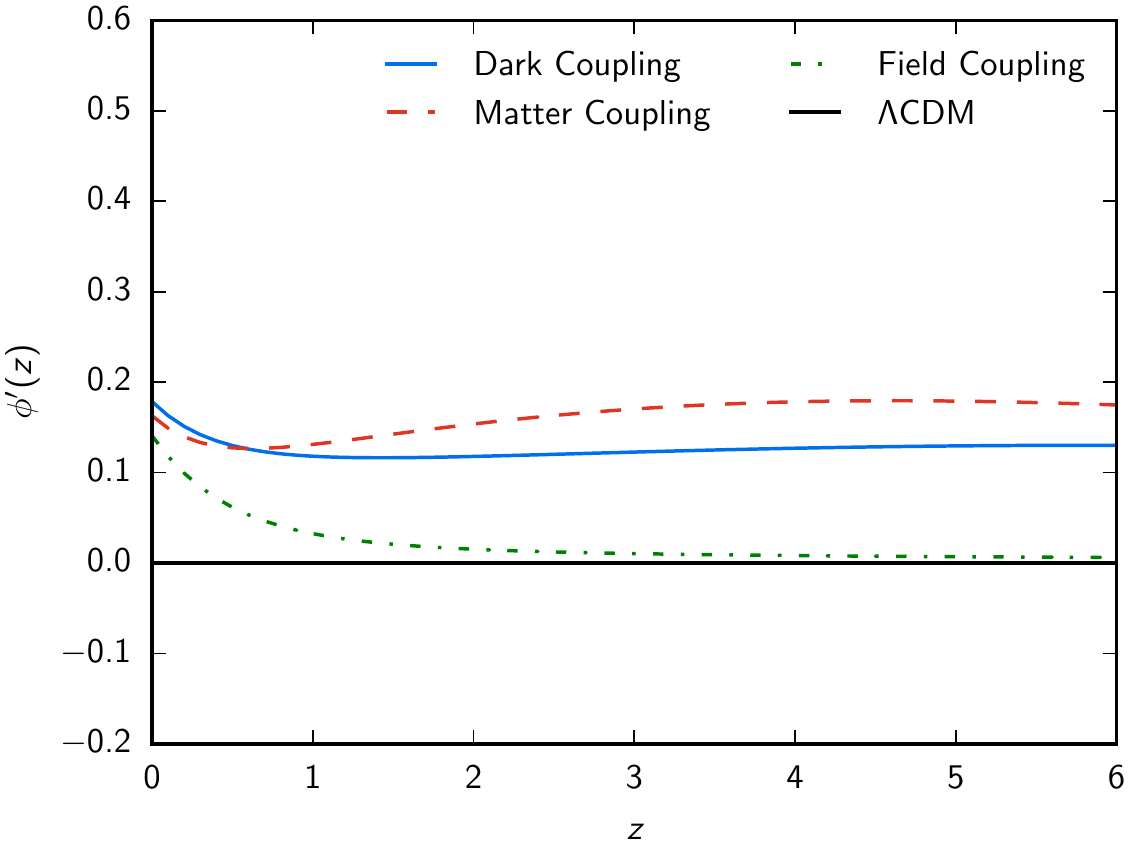}
\caption{Redshift evolution of the dilaton scalar field $\phi$ (left panel) and of its derivative $\phi'$ (right panel)
in the Dark (blue solid line), Matter (red dashed line) and Field (green dot-dashed line) Coupling cases, assuming $\alpha_{b,0}=10^{-4}$ and $\alpha_V=0.1$. These quantities 
vanish in the standard $\Lambda$CDM scenario, as shown with the black solid line for reference.}
\label{fig:field}
\end{center}
\end{figure*}

We show the evolution of the runaway dilaton field and of its derivative in the 3 different coupling
choices in Figure \ref{fig:field}. 
The curves are generated using the Planck 2015 marginalized values of the cosmological parameters \cite{Ade:2015xua} 
and assuming the baryonic and dark energy couplings to be $\alpha_{b,0}=10^{-4}$ and $\alpha_V=0.1$.

\subsection{Effects on cosmological observables} 
\label{sec:obs}
For our baseline analysis we compute 3 main observables: 
\begin{enumerate}
\item the redshift dependence of the fine structure constant $\alpha$: 
\begin{equation}\label{eq:dalpha}
\frac{\Delta\alpha}{\alpha}(z)\equiv\frac{\alpha(z)-\alpha_0}{\alpha_0}=B_F^{-1}(\phi)-1=\frac{\alpha_b}{40}
\left[1-e^{-(\phi(z)-\phi_0)}\right] \,.
\end{equation}
As discussed in \cite{Martins:2015dqa}, the time variation of $\alpha$ is an immediate consequence of the coupling of the dilaton with the electromagnetic sector of the theory.
\item the redshift evolution for the commonly used Type Ia Supernovae reduced magnitude $\mu(z)$, relating the Supernovae's apparent, $m$, and absolute, $M$, magnitude to the universe 
expansion:
\begin{equation}
 \mu(z)\equiv m-M=5\log_{10}{\left[(1+z)\frac{c}{H_0}\int_0^z{\frac{dz'}{E(z')}}\right]}+25\,.
\end{equation}
This can be measured with the E-ELT through Supernovae up to redshifts $z\approx4$ \cite{Hook} (as further described below).
\item the change in the spectroscopic velocity of distant sources $\Delta v$ 
due to the so-called redshift drift phenomenon \cite{Sandage,Loeb:1998bu}:
\begin{equation}
 \Delta v = cH_0 \Delta t \Big[ 1-\frac{E(z)}{1+z}\Big]\,,
\end{equation}
where $\Delta t$ is the time interval between two observations of the same astrophysical source.
This has been shown to be relevant for the investigation of dark energy theories 
\cite{Liske:2008ph, Corasaniti:2007bg, Quercellini:2010zr,Martinelli:2012vq,Calabrese:2013lga}, in particular thanks to the wide range
of redshift at which it can in principle be detected (up to $z\approx5$). 
\end{enumerate}

\begin{figure*}[t]
\begin{center}
\includegraphics[width=10.5cm]{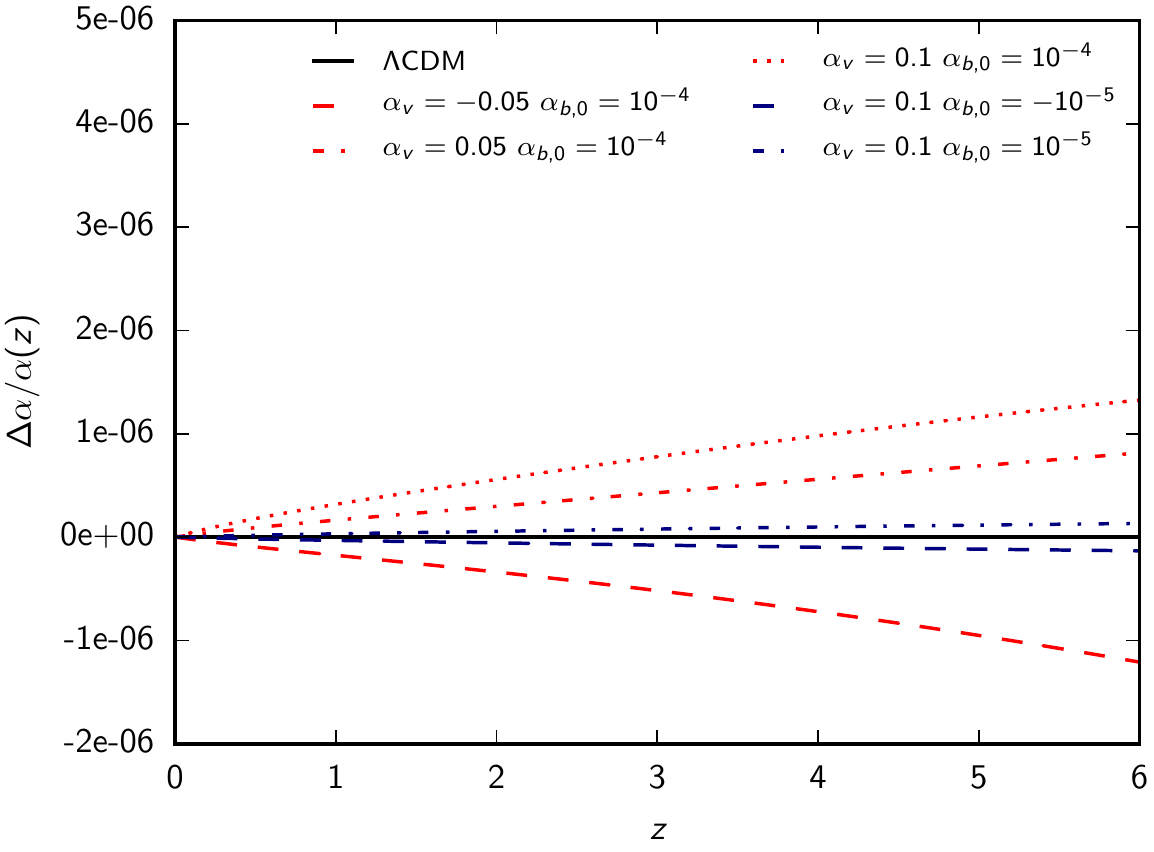}
\caption{Redshift dependence of the relative variation of $\alpha$ in the Dark Coupling case for several values of the coupling parameters. A similar behaviour is observed if we consider 
the Matter and Field coupling cases, although with different amplitudes. We note that the departures from the standard model increase for higher values of the couplings. We anticipate that 
this effect will be probed by future data (with a sensitivity of e.g., 
$\sigma_{\alpha}\sim10^{-7}$).}
\label{fig:dalpha}
\end{center}
\end{figure*}

Figure \ref{fig:dalpha} shows the evolution of $\alpha$ with redshift for some values of the coupling 
parameters $\alpha_{b,0}$ and $\alpha_V$, and highlights a strong dependence of the $\alpha$ variation on the amplitudes of the couplings. Measurements of the fine structure constant are 
then extremely promising to set constraints on dilaton scenarios. We note that in this class of models the two different particle physics parameters determine the evolution on $\alpha$ in 
different ways: $\alpha_{b,0}$ provides the overall normalization, while $\alpha_V$ is driving the evolution of the field itself. On the other hand, these two parameters play different 
roles in the behaviour of dark energy (where the latter one has the dominant effect); for this reason, this is a Class II model in the terminology of \cite{GRG}.\\

\begin{figure}[!t]
\begin{center}
\includegraphics[width=10.5cm]{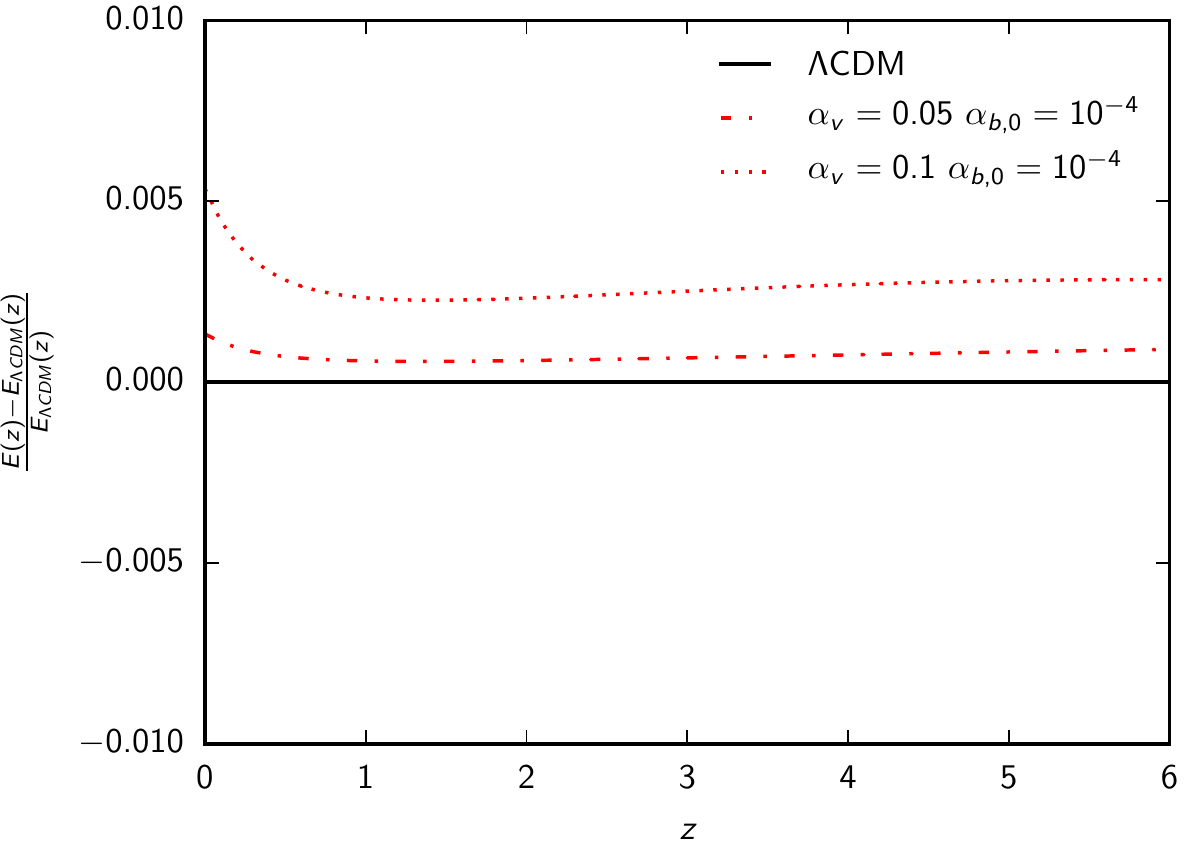}
\caption{Relative difference between the $E(z)$ predicted in the runaway dilaton scenario, for different choices of the parameters, and the one predicted by $\Lambda$CDM. The departures from 
the standard expansion are at the sub-percent level at all redshifts and so hard to detect even with future data.}
\label{fig:backev}
\end{center}
\end{figure}

We noted before that the potential of the dilaton affects the expansion of the universe as an effective cosmological constant. However, Eq.(\ref{eq:fried1}), re-expressed as:
\begin{equation}
E(z)\equiv\frac{H(z)}{H_0}=\frac{\sqrt{\Omega_m(1+z)^3+\Omega_V}}{\sqrt{1-\phi'^2/3}} \,,
\end{equation}
shows that, given our assumption of a $\Lambda$-like effective Dark Energy, deviations from the standard $\Lambda$CDM background behaviour are only generated 
by the contribution of the field kinetic energy. Nevertheless, the current constraints on $|\phi'_0|$ and on the couplings $\alpha_{b,m,V}$ imply that $\phi'$ is small at low redshifts. This 
leads to small departures from the standard $E(z)$ even when the couplings are non-vanishing, which reflects in an evolution of the background observables very 
similar to the standard case. 
In Figure \ref{fig:backev} we show how the departures from the $\Lambda$CDM predictions brought to the $E(z)$ function by the dilaton are well below $1\%$ 
at all redshifts considered. We, therefore, anticipate that the effect of the dilaton cannot be resolved at high significance with background expansion observables, 
at least with the data expected in the upcoming years\footnote{This result disagrees with Figure 4 of \cite{Martins:2015dqa} where the dilaton field brought 
to a significant impact on the redshift drift  signal; however the result shown there was affected by a bug in the calculation in the field derivative $\phi'(z)$.}. We note that larger 
differences may arise if the $w=-1$ assumption on the effective Dark Energy is relaxed.\\

Although not investigated in this paper --which focuses primarily on the E-ELT sensitivity-- the coupling of the dilaton field with the electrodynamic sector of the theory might also lead to 
other observable effects potentially useful to constrain this kind of models. 
As it is generically the case for scalar fields coupled with the electromagnetic tensor \cite{AMMVL}, the dilaton might lead for example to a non-standard propagation of CMB photons.

\section{Simulated Data and Analysis}
\label{sec:ana}

In this paper we focus on future observations from the upcoming European
Extremely Large Telescope (E-ELT), whose first light is expected in 2024. This new generation 39.3 meter telescope will be equipped with an ultra-stable, high-precision 
spectrograph (ELT-HIRES, whose top-level requirements are described in \cite{HIREStlr}) which, via spectroscopic observations of absorption systems along the line of sight of bright quasars, 
will be able to provide measurements of the fine structure constant with a sensitivity of $\sigma_\alpha=10^{-7}$.

We simulate 3 datasets of $\Delta\alpha/\alpha$ measurements provided by E-ELT assuming the 
observation of 100 QSO absorption systems uniformly distributed in the redshift range $0.5<z<4$. We fix the standard cosmological parameters to the latest CMB Planck 2015 marginalized values 
\cite{Ade:2015xua} and we consider 3 choices of the dilaton parameters:
\begin{itemize}
 \item $\alpha_{b,0}=0\,, \alpha_V=0$
 \item $\alpha_{b,0}=1\times10^{-5}\,, \alpha_V=0.05$
 \item $\alpha_{b,0}=5\times10^{-5}\,, \alpha_V=0.1$
\end{itemize}
where the last two cases are generated assuming the Dark Coupling relation between $\alpha_m$ and $\alpha_V$.

The values of these couplings have been chosen to exploit how the constraints change moving from a $\Lambda$CDM cosmology to more and more extreme, but still allowed by current
constraints, departures from the standard paradigm.
The first case recovers the $\Lambda$CDM model and will inform us on how well we can reduce the parameters space of the dilaton model if data are in agreement with the standard model. The  
last two cases (the non-vanishing $\alpha_i$) will investigate the sensitivity of future data to the dilaton parameters if, and at which degree, the data manifest departures from $\Lambda$CDM.

Despite expecting little additional information from background observables (as we discussed in Section \ref{sec:obs}), we simulate and include in our analysis Supernovae and redshift drift 
data as expected from E-ELT.
This will help in breaking degeneracies between parameters and provide better constraints on matter-dark energy densities and the Hubble expansion rate. 
Furthermore, by including these background observables we will perform a complete analysis for expected E-ELT cosmological probes and then assess the possibility of constraining 
the dilaton model with this facility alone, without having to rely on synergies with other experiments (with one caveat that we now discuss).

The James Webb Space Telescope (JWST, through NIRcam imaging), should find about 50 type Ia Supernovae and measure their light curves \cite{RiessLivio}, and with E-ELT spectroscopy provided 
by HARMONI \cite{Thatte} the redshift and Supernova type can be confirmed. The redshift range of this high-z sample is expected to be $1<z<4$. The redshift distribution of these Supernovae is 
not easy to extrapolate, since even the most detailed current studies such as those of the SNLS team \cite{Perrett} only reach out to $z\sim1$. In the absence of a 
specific redshift distribution, we will simply assume it to be uniform in the above range.

Given its cutting edge stability and repeatability, the ELT-HIRES is suited to investigate the redshift drift signal by observing the Lyman $\alpha$ absorption lines of distant QSOs as 
discussed in \cite{Liske:2008ph}, which shows 
how the E-ELT will be able to detect the redshift drift with a 4000 hours of integration in a period of $\Delta t=20$ 
years. (This time may be significantly reduced if additional bright targets are found.) Monte Carlo simulations performed in the context of the COsmic Dynamics Experiment (CODEX) 
Phase A study \cite{codex}, 
predict an error on the spectroscopic velocity shift $\Delta v$ that can be expressed as:
\begin{equation}\label{eq:error}
 \sigma_{\Delta v}=1.35\ \frac{2370}{S/N}\ \sqrt{\frac{30}{N_{\rm QSO}}}\ \left(\frac{5}{1+z_{\rm QSO}}\right)^x\ \rm{cm\ s^{-1}},
\end{equation}
where $S/N$ is the signal to noise ratio, $N_{\rm QSO}$ the number of observed quasars, $z_{\rm QSO}$ their redshift 
and the exponent $x$ is $1.7$ for $z\leq4$ and $0.9$ for higher redshifts.

Using these specifications we produce a redshift drift mock dataset with $S/N=3000$ 
and $N_{\rm QSO}=30$ assumed to be uniformly distributed among the following redshift bins $z_{\rm QSO}=[2.0,2.8,3.5,4.2,5.0]$. As for $\alpha$, we fix the standard 
cosmological parameters to the Planck 2015 values. 

We analyze the E-ELT data using the publicly available code \texttt{COSMOMC} \cite{Lewis:2002ah}, modified to generate the theoretical predictions of the reduced magnitude, the 
redshift drift   and the variation of the fine structure constant in the runaway dilaton model for any value of the free parameters of the theory. 
We sample five parameters: the baryon and cold dark matter densities $\Omega_{b}h^2$ and $\Omega_{c}h^2$, the ratio between the sound horizon and the angular diameter distance at 
decoupling $\theta_s$, and the runaway dilaton free parameters $\alpha_{b,0}$ and $\alpha_V$.
We assume flat priors on all the parameters and we bound the variation of the dilaton couplings using the currently available constraints described in Section \ref{sec:theo}, 
i.e. $|\alpha_{b,0}|\leq10^{-4}$ and $|\alpha_V|\leq0.4$.
However, when we report the results of our analysis (e.g., in Table \ref{tab:results},\ref{tab:nonLCDM}) we quote the constraints obtained on the derived cosmological parameters --matter 
density and 
Hubble parameter-- and on the dilaton parameters. We do not report individual limits on baryonic and dark matter densities $\Omega_bh^2$ and $\Omega_ch^2$ 
as the cosmological probes considered here are not sensitive to them separately, but rather to the total matter density $\Omega_m=\Omega_b+\Omega_c$.\\

The simultaneous variation of cosmological and dilaton parameters allows us to investigate both the degeneracies between them and the internal degeneracy of the dilaton couplings,
$\alpha_b$ and $\alpha_V$; given the competing effects of these 2 parameters on E-ELT observables, we can expect a strong impact of this degeneracy on the constraining power of this
upcoming facility. Moreover, the exploration of 3 physical models for the evolution of the dark matter coupling $\alpha_m$ allows us to test whether or not E-ELT will be able to distinguish 
between different models of the same runaway dilaton family.\\

\section{Results}
\label{sec:res}

Our analysis shows that, depending on the values of the dilaton couplings, there are three different regimes for which the E-ELT data will have different impacts. We will refer to these as 
the null case (vanishing couplings), the weak coupling case (where the couplings, although non-zero, can't be statistically distinguished from zero due to degeneracies) and the strong 
coupling case (where the non-zero couplings are compatible with currently existing bounds but can be detected by the E-ELT). We now illustrate each of these cases. We also recall that, 
although the simulated datasets are generated assuming Dark Coupling for $\alpha_m$, for each case we will have three possible scenarios according to the theoretical choice of the dark matter 
coupling made in the analysis. 

\subsection{Null case}

\begin{table}[!t]
\begin{center}
\begin{tabular}{l|c|c|c|c}

                         & Fiducial    & Dark Coupling                   & Matter Coupling                    & Field Coupling \\
\hline
$\Omega_m$               & $0.314$     & $0.309\pm 0.011$                & $0.308\pm 0.012$                   & $0.310\pm 0.011$   \\
$\alpha_{b,0}$           & $0.0$       & $(0.0\pm 2.5)\times10^{-5}$     & $(0.1^{+1.2}_{-1.4})\times10^{-5}$ & $(0.0\pm 3.0)\times10^{-5}$  \\
$\alpha_V$               & $0.0$       & $0.035^{+0.04}_{-0.10}$         & $0.038^{+0.05}_{-0.11}$            & $0.009^{+0.08}_{-0.10}$  \\
$H_0$                    & $67.26$     & $67.4^{+1.4}_{-1.0}$            & $67.4^{+1.5}_{-1.0}$               & $67.6^{+1.1}_{-1.0}$    \\
$\phi'_0$                & $0$         & $-0.061^{+0.19}_{-0.08}$        & $-0.062^{+0.19}_{-0.09}$           & $-0.01^{+0.15}_{-0.13}$   \\
\hline
\end{tabular}
\caption{Marginalized values and $68\%$ c.l. limits for the analyses of the null case. We report the fiducial values of the parameters used to simulate the data as a reference in 
column 1. The following 3 columns separate the results by different theoretical coupling assumptions made in the analysis (see Section \ref{sec:theo}).}
\label{tab:results}
\end{center}
\end{table}

\begin{figure}[h]
 \centering
 \includegraphics[width=7.65cm]{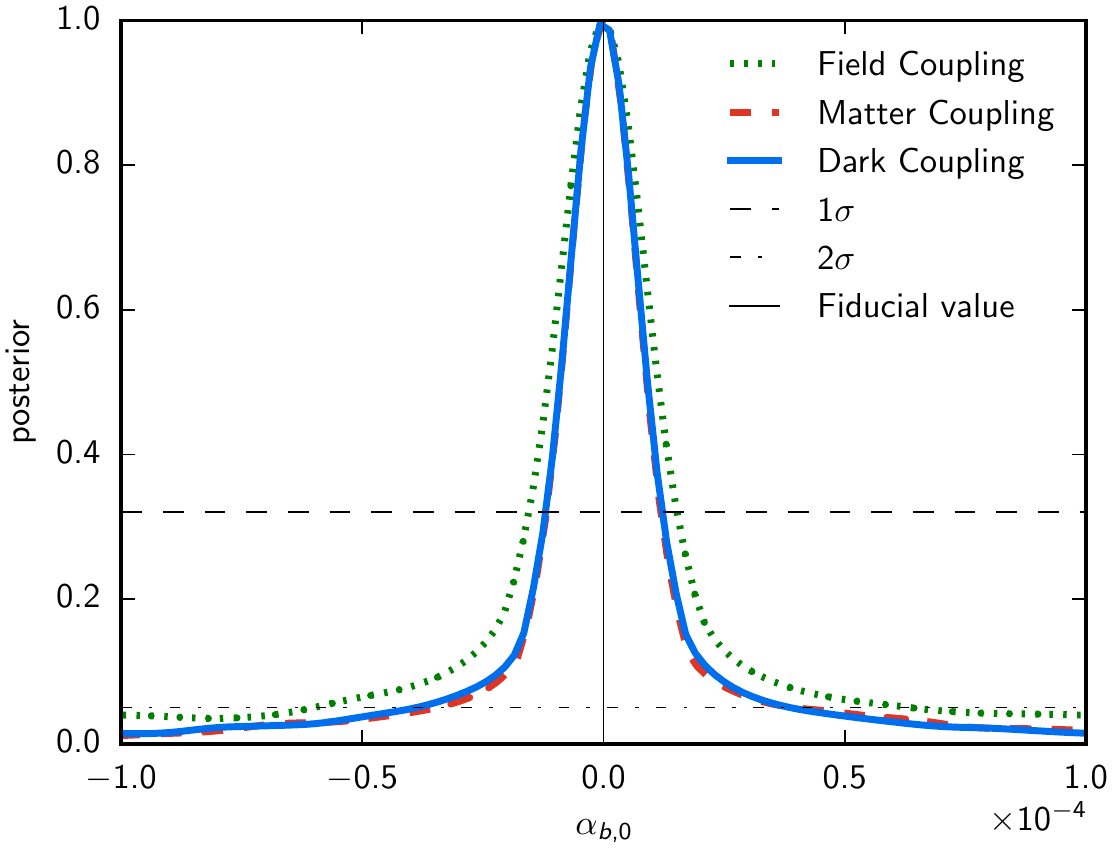}
 \centering
 \includegraphics[width=7.65cm]{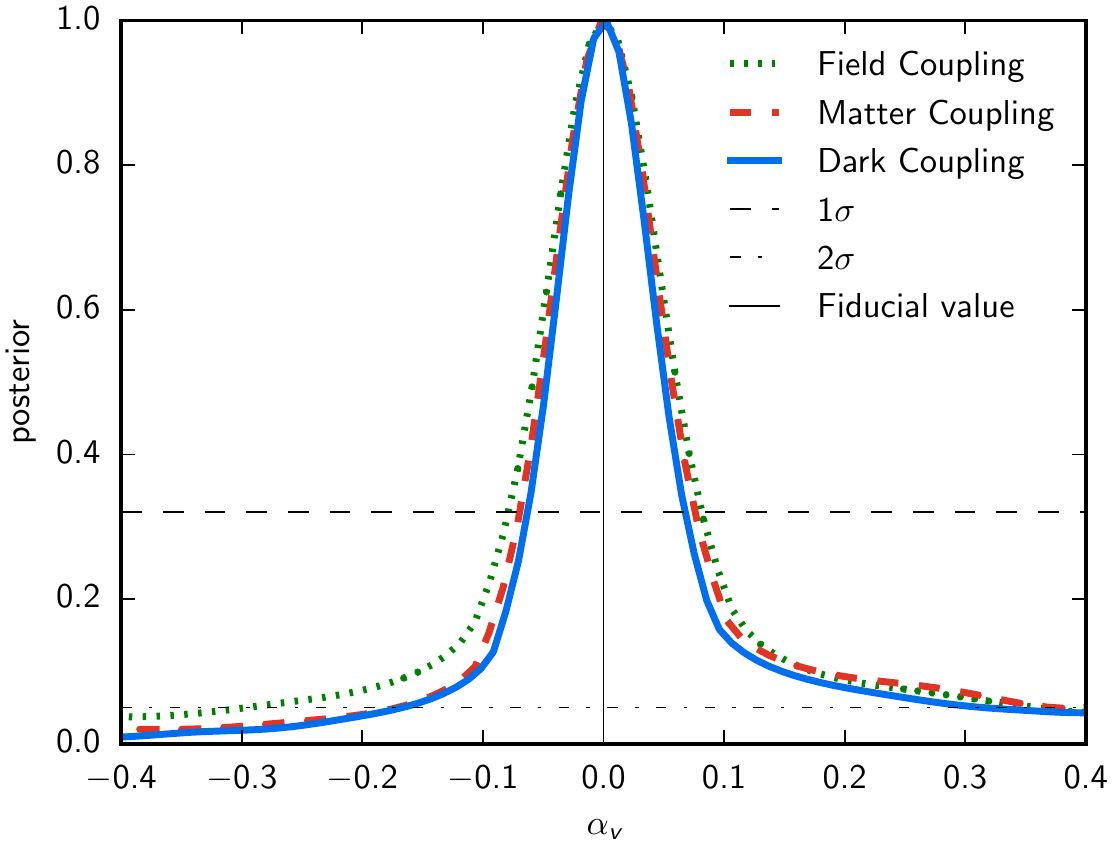}
 \centering
\includegraphics[width=7.65cm]{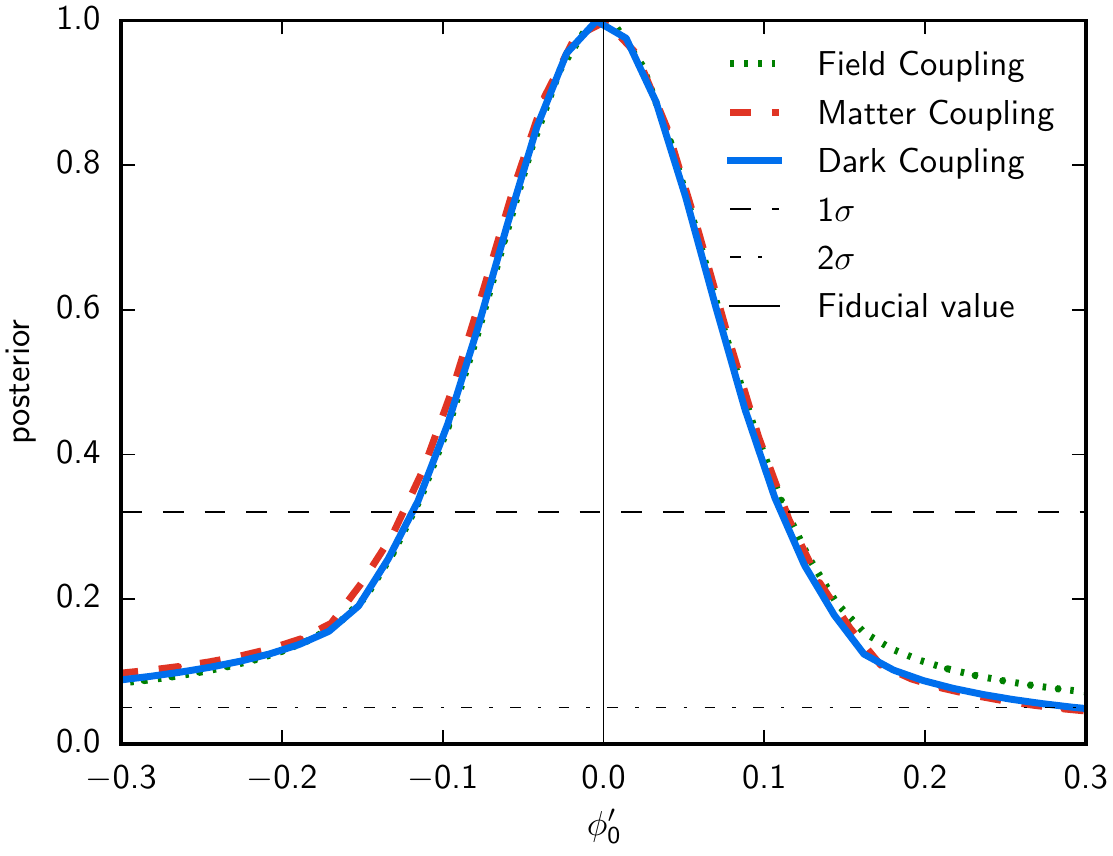}
\caption{Posterior distributions for the dilaton couplings $\alpha_{b,0}$ (top left panel) and $\alpha_V$ (top right panel) and for the potential derivative $\phi'_0$ (bottom panel)
derived from Eq.(\ref{eq:incond}). These posteriors are obtained analyzing the null case in the 3 coupling assumptions.}
\label{fig:post}
\end{figure}

For the null case (i.e., input vanishing couplings reproducing a $\Lambda$CDM universe), reported in Table \ref{tab:results} we notice that, in spite of the high sensitivity of the E-ELT on 
$\alpha$ variations, the constraints that it will provide on the coupling parameters and, consequently, on $\phi'_0$ only improve upon the prior range we assumed following Section \ref{sec:obs} 
by a factor of a few (see also Figure \ref{fig:post}). 
However, Figure \ref{fig:post} highlights how the $\alpha_V$ posterior excludes more, at the 2-$\sigma$ level, the negative tail. This is a consequence of the fact that the variation of 
$\alpha$ deviates more from the $\Lambda$CDM expectation when negative $\alpha_V$ are considered, as can be seen in Figure \ref{fig:dalpha} for the $\alpha_V=0.05$ and $\alpha_V=-0.05$ 
predictions.

\begin{figure}[!t]
\begin{center}
\hspace*{-1.5cm}
\begin{tabular}{cc}
\includegraphics[width=10cm]{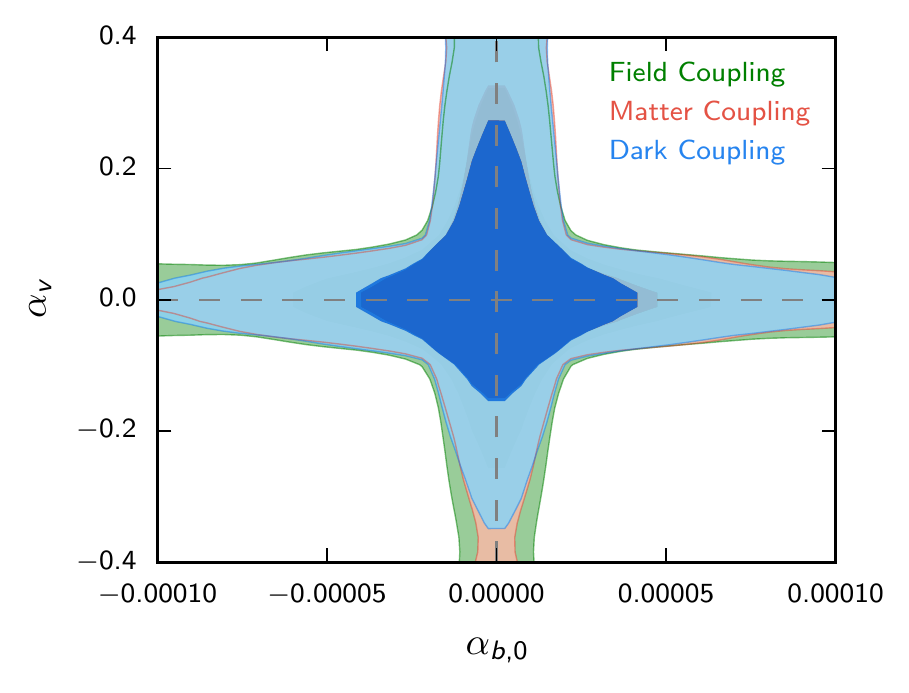} 
 \end{tabular}
\caption{68\% and 95\% c.l. contours in the $\alpha_{b,0}$-$\alpha_V$ plane for null case analyses in the 3 coupling assumptions. The grey dashed lines 
identify the fiducial values.}
\label{fig:abav}
\end{center}
\end{figure}

To understand this relatively mild improvement provided by E-ELT one needs to further explore the degeneracy between the coupling parameters.
Figure \ref{fig:abav} shows the 2-Dimensional $1$ and 2-$\sigma$ contours in the plane $\alpha_{b,0}$-$\alpha_V$, highlighting how a strong degeneracy arises when a 
$\Lambda$CDM model is used as fiducial cosmology.
This behaviour is connected to the dependence of the variation of the fine structure constant on the coupling parameters, indeed setting one of the couplings to zero will 
reproduce exactly the standard non varying fine structure constant. Moreover, both Figure \ref{fig:abav} and Table \ref{tab:results} show that the 3 coupling cases are 
undistinguishable as they all reduce to the $\Lambda$CDM scenario in the same way.

\begin{figure*}[!ht]
\begin{center}
\includegraphics[width=7.65cm]{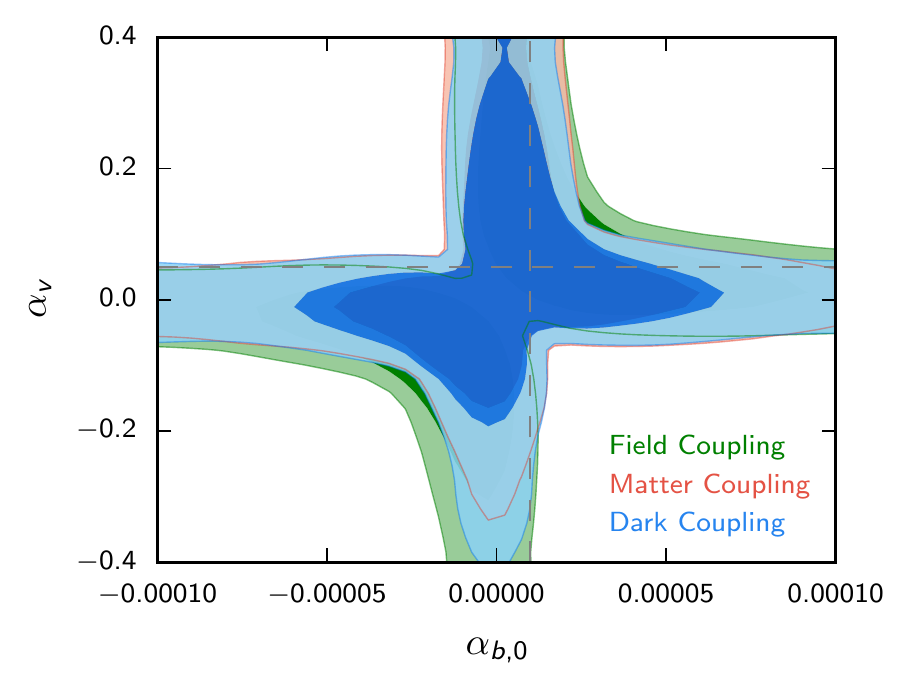}  
\includegraphics[width=7.65cm]{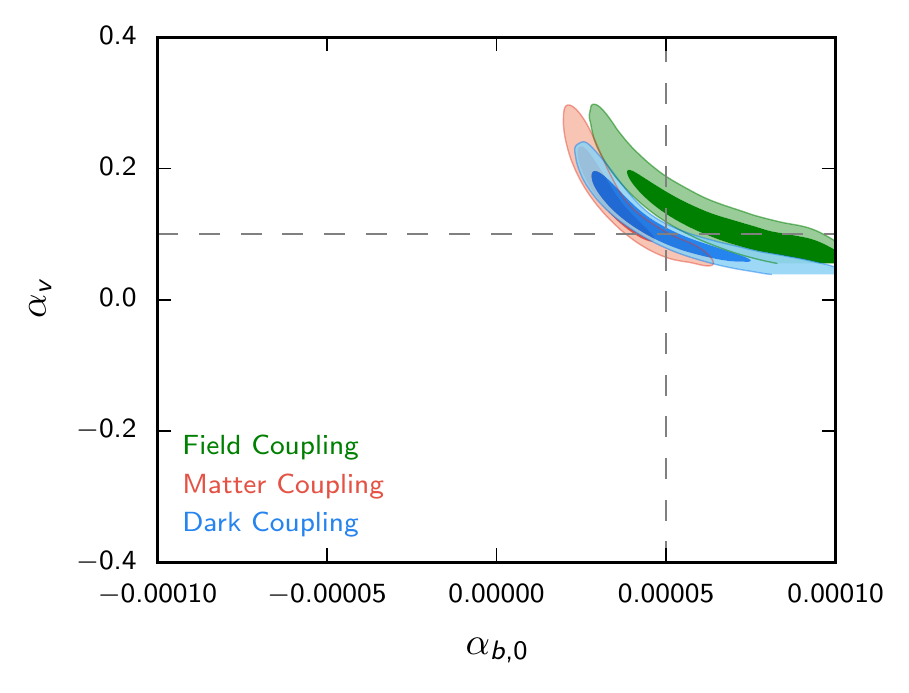}
\caption{68\% and 95\% contours in the $\alpha_{b,0}$-$\alpha_V$ plane for weak (left panel) and strong (right panel) coupling analyses in the 3 coupling assumptions. 
The grey dashed lines identify the fiducial values.}
\label{fig:nonLCDM}
\end{center}
\end{figure*}

\subsection{Weak coupling case}

\begin{figure*}[!t]
\begin{center}
\includegraphics[width=7.65cm]{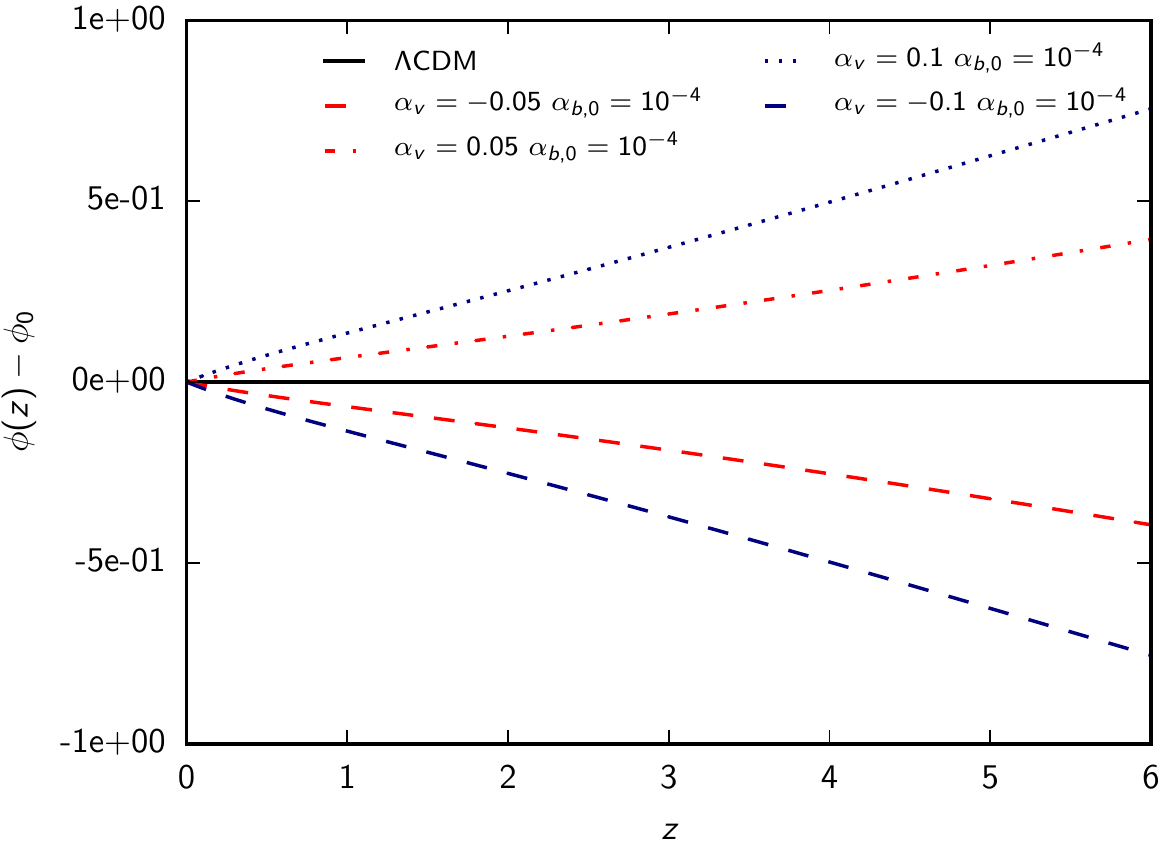}  
\includegraphics[width=7.65cm]{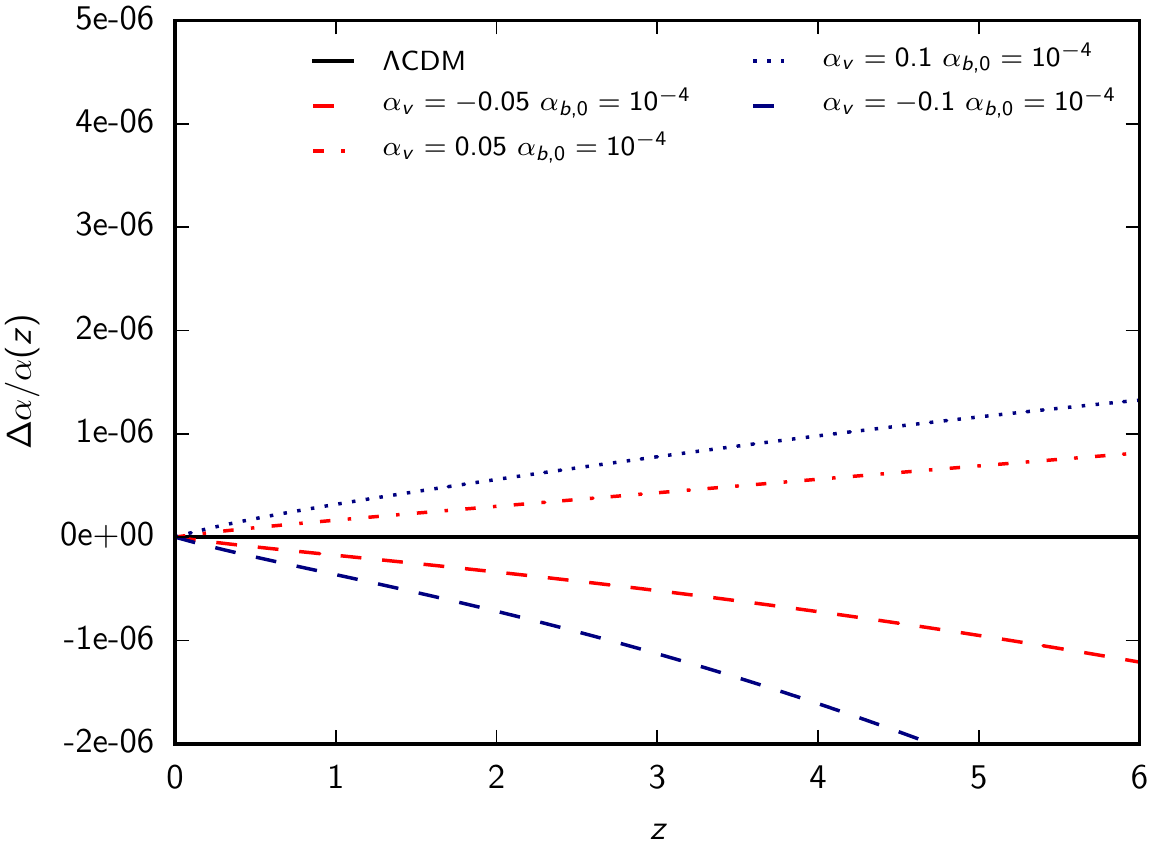}
\caption{Field (left panel) and $\Delta\alpha/\alpha$ (right panel) dependence on the $\alpha_V$ parameter in the Dark Coupling case. While $\phi(z)$ is anti-symmetric
with respect to $\alpha_V=0$, the variation of $\alpha$ preserves this symmetry only for small values of $\alpha_V$.}
\label{fig:simm}
\end{center}
\end{figure*}

The degeneracy of these two coupling parameters can be further characterized if we analyze the case of non-vanishing but small couplings described in Section \ref{sec:ana}. The 
left panel of Figure \ref{fig:nonLCDM} shows the 1 and 2-$\sigma$ 2D contours obtained for the fiducial $\alpha_{b,0}=10^{-5}$ and $\alpha_V=0.05$. In this case the degeneracy is still 
present but not completely symmetric as before; negative values of the couplings (away from the positive fiducial values) can still mimic a variation of $\alpha$ 
compatible with the simulated dataset. 
This behaviour can be easily explained noticing that the overall sign of the $\alpha$ variation in Eq.(\ref{eq:dalpha}) is given by $\alpha_{b,0}$ and that the same 
equation can be approximated for $\phi(z)-\phi_0<<1$ as:
\begin{equation}\label{eq:approx}
 \frac{\Delta\alpha}{\alpha}(z)\approx\frac{\alpha_{b,0}}{40}\left[\phi(z)-\phi_0\right];
\end{equation}
the left panel of Figure \ref{fig:simm} shows that $\phi(\alpha_V)\approx-\phi(-\alpha_V)$ and therefore, in the approximation of small field values
\begin{equation}
 \frac{\Delta\alpha}{\alpha}(\alpha_V)\approx-\frac{\Delta\alpha}{\alpha}(-\alpha_V)\Rightarrow
 \frac{\Delta\alpha}{\alpha}(\alpha_{b,0},\alpha_V)\approx\frac{\Delta\alpha}{\alpha}(-\alpha_{b,0},-\alpha_V).
\end{equation}

The degeneracy between $\alpha_{b,0}$ and $\alpha_V$ is therefore the physical bottleneck preventing the extremely sensitive E-ELT data from constraining these parameters with high 
significance; this means 
that a future line of investigation, outside the aim of the present paper, relies on the identification of other observational signatures of the runaway
dilaton model able to break this degeneracy and therefore to significantly tighten the constraints. Improved local constraints on Weak Equivalence Principle violations will 
also help, by providing tighter priors on the couplings.

\subsection{Strong coupling case}

Last but not least, we show how the $\alpha_{b,0}$-$\alpha_V$ plane changes significantly when we consider the third set of simulated datasets, $\alpha_{b,0}=5\times10^{-5}$ and $\alpha_V=0.1$.
These are compatible with (but not far from) the current bounds, as we discussed earlier.

For values of $\alpha_V\gtrsim0.05$ the approximation of Eq.(\ref{eq:approx}) breaks down (see also Figure \ref{fig:simm}); 
this leads to the behaviour observed in the right panel of Figure \ref{fig:nonLCDM}, where the fiducial model can be recovered only with the right sign of the couplings. This result 
shows how moving away from the $\Lambda$CDM fiducial model allows to obtain stronger constraints on the parameters. When the approximation of Eq.(\ref{eq:approx}) breaks down, 
the $\Delta\alpha$  variation is more sensitive to the coupling and the parameters can be better determined, as shown in Table \ref{tab:nonLCDM}.

\begin{table}[!t]
\begin{center}
\begin{tabular}{l|c|c|c|c}

                 & Fiducial             & Dark Coupling                        & Matter Coupling                      & Field Coupling               \\
\hline
$\Omega_m$       & $0.314$              & $0.309\pm 0.011$                     & $0.307\pm 0.011$                     & $0.310\pm 0.011$             \\
$\alpha_{b,0}$   & $5.0\times10^{-5}$   & $(4.94^{+0.8}_{-2.2})\times10^{-5}$  & $(3.48^{+0.4}_{-1.0})\times10^{-5}$  & $(6.6\pm 1.9)\times10^{-5}$  \\
$\alpha_V$       & $0.1$                & $0.116^{+0.03}_{-0.06}$              & $0.158^{+0.05}_{-0.06}$              & $0.126^{+0.02}_{-0.06}$      \\
$H_0$            & $67.26$              & $67.7\pm 1.0$                        & $67.6\pm 1.0$                        & $67.7\pm 1.0$                \\
$\phi'_0$        & $-0.20$              & $-0.204^{+0.11}_{-0.05}$             & $-0.259^{+0.09}_{-0.07}$             & $-0.192^{+0.09}_{-0.03}$     \\
\hline
\end{tabular}
\caption{Marginalized values with their $68\%$ c.l. limits obtained analyzing the strong coupling case. The columns follow the same notation of Table 
\ref{tab:results}. We recall that the strong case data are simulated assuming Dark Coupling for $\alpha_m$.}
\label{tab:nonLCDM}
\end{center}
\end{table}

\begin{figure*}[!t]
\begin{center}
\includegraphics[width=7.65cm]{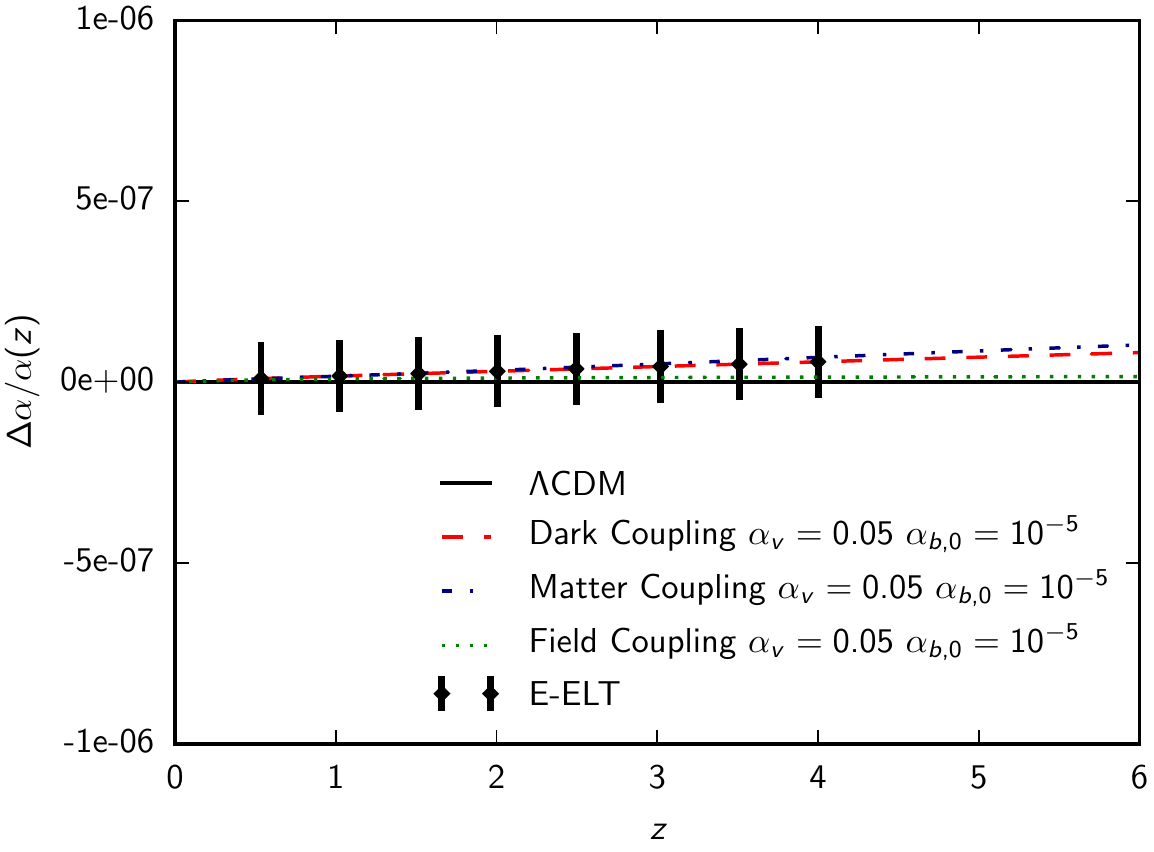} 
\includegraphics[width=7.65cm]{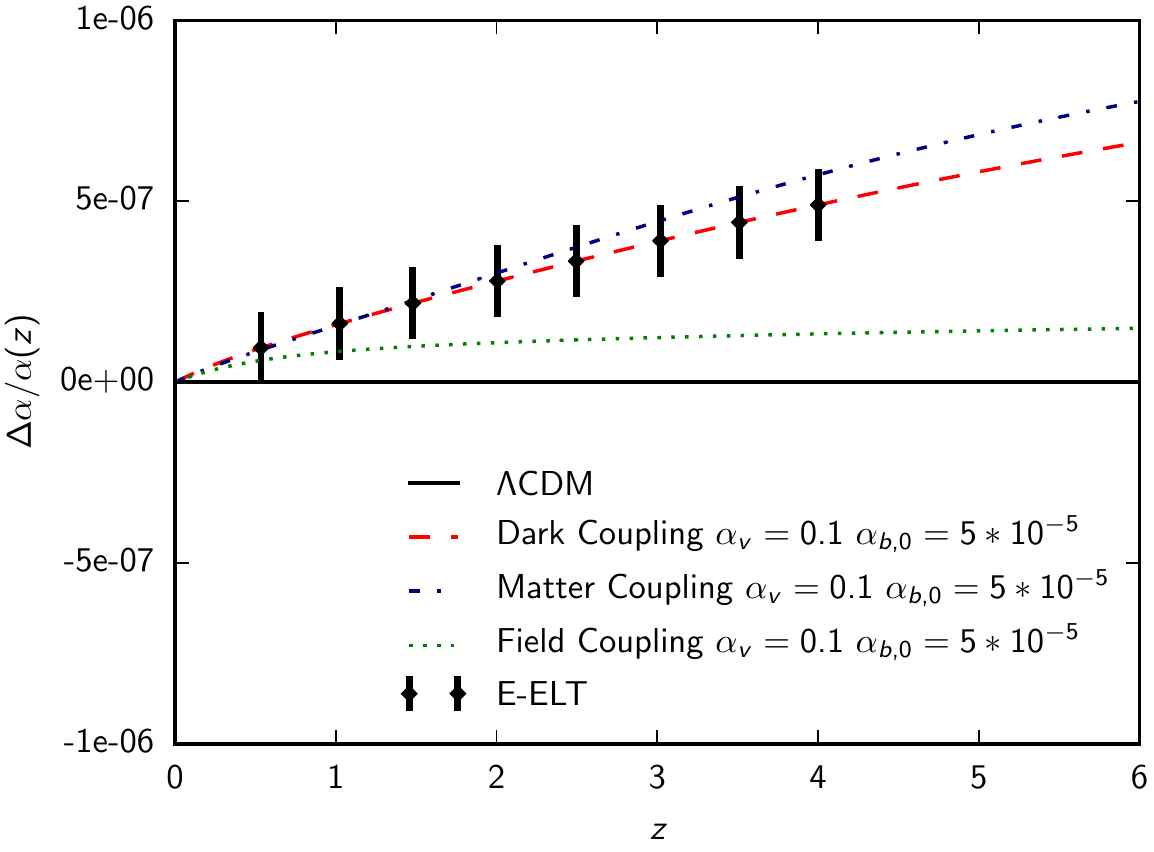}
\caption{Simulated E-ELT data (black dots and error bars) in the two non-standard fiducial cosmologies assumed in Section \ref{sec:ana}  (for plotting reasons only a few of the dataset 
points are shown in the plot). It is possible to notice how 
in the most extreme case (right panel), with $\alpha_{b,0}=5\times10^{-5}$ and $\alpha_V=0.1$, the theoretical expectations in the 
3 coupling cases considered are not able to fit the data for the same values of the parameters, while the $\alpha_{b,0}=1\times10^{-5}$ and $\alpha_V=0.05$ dataset (left panel) is
simultaneously well fitted by all 3 choices of $\alpha_m$.}
\label{fig:fiducial}
\end{center}
\end{figure*}

Moreover, in this extreme fiducial cosmology, the physical difference between the 3 choices for $\alpha_m$ starts to emerge. This effect arises from the fact that while the 3 models produce 
similar behaviours of the fine structure constant for values of the parameters close to the $\Lambda$CDM limit (i.e., in the weak coupling limit), the differences between them increase 
moving away from it and in the strong case they become distinguishable. In the Field Coupling case, for instance, Eq. (\ref{eq:field}) loses the source term coupled by $\alpha_m$ and 
the net effect is a smaller amplitude of the field $\phi$ (see Fig. \ref{fig:field}); this implies that the Field Coupling produces smaller variations of the fine structure constant,
 as can be seen in Fig. \ref{fig:nonLCDM}. 
These differences among the 3 investigated coupling choices lead to the impossibility of correctly fitting the simulated dataset with any $\alpha_m$ assumption. Figure \ref{fig:nonLCDM} 
shows that the input model, which assumes the Dark Coupling mechanism, cannot be reproduced correctly with the Field Coupling mechanism, but is still partially in agreement with the Matter 
Coupling case (see also the degenerate curves shown in Figure \ref{fig:fiducial}).\\
This however does not translate in an observable bias on the standard cosmological parameters which could be 
used in principle, when real data will be available, as a tool to establish which, if any, of the coupling mechanisms is driving the variation of $\alpha$.
To observe this kind of bias we need to identify different observational probes which are more connected to the standard cosmological parameters and can therefore probe the 
degeneracies of these with the dilaton couplings.

\section{Conclusions}
\label{sec:conc}

In this paper we investigated the possibility of constraining the string inspired runaway dilaton model by exploiting the upcoming observational data from some of E-ELT's suite of instruments. 
Specifically, ELT-HIRES will improve the sensitivity of measurements of the fine structure constant $\alpha$ (and map its possible redshift variation) and measure the redshift drift  of objects 
following the Hubble flow, while HARMONI will complement other cosmological datasets by characterizing high-redshift Type Ia Supernovae.

We have shown how the runaway dilaton model predicts a variation of $\alpha$ with an amplitude and redshift dependence directly connected to the evolution of the dilaton 
field $\phi(z)$ and therefore to its couplings to baryonic matter, dark matter and effective dark energy. We then simulated $\Delta\alpha$ measurements as expected from E-ELT to 
investigate its sensitivity to the dilaton model parameters, alongside with redshift drift and Supernovae measurements from the same facility, used primarily to constrain standard background 
cosmological parameters. 

Assuming a $\Lambda$CDM fiducial cosmology, we confirmed that the E-ELT measurements will improve constraints on $\alpha_V$ with respect to those currently available (by a factor of order 
a few), while constraints on the baryonic coupling $\alpha_{b,0}$ will not be competitive with the ones expected from forthcoming local equivalence principle
tests; the main reason behind this is the strong degeneracy between the two parameters that arises in $\Delta\alpha$ measurements, which still plagues the results even if the fiducial model 
is slightly shifted away from the standard $\Lambda$CDM cosmology. It will be interesting in the future to investigate other observable effects of the runaway dilaton model; some of these, 
like a violation of the CMB temperature-redshift relation, can also be constrained by 
ELT-HIRES \cite{HIREStlr}.

On the other hand there is a significant region of the parameters space that is compatible with all currently available cosmological and local (laboratory) constraints for which the E-ELT 
would be able to detect deviations from the standard model at a good level of significance, since in this regime the degeneracy between $\alpha_V$ and $\alpha_{b,0}$ is partially
broken.

We conclude that E-ELT will be a crucial facility to investigate the runaway dilaton model, but it can obtain strong constraints on the model's parameters only when 
the cosmological model departs significantly from the standard $\Lambda$CDM; broadly speaking, its sensitivity is optimal for couplings within one order of magnitude of the currently 
available bounds. Should this not be the case, other observable effects of the dilaton need to be investigated as measurements of $\Delta\alpha$ alone will not allow significantly improved 
constraints due to the parameter degeneracies we have highlighted. In any case, astrophysical tests carried out by the E-ELT will provide an important complement to local equivalence 
principle tests.

\section*{ACKNOWLEDGMENTS}

This work was supported by Funda\c c\~ao para a Ci\^encia e a Tecnologia (FCT) through the research grants PTDC/FIS/111725/2009 and UID/FIS/04434/2013. MM acknowledges the DFG TransRegio 
TRR33 grant on The Dark Universe. EC is supported by a Beecroft Fellowship and the STFC. CJM is supported by an FCT Research Professorship, 
contract reference IF/00064/2012, funded by FCT/MCTES (Portugal) and POPH/FSE (EC).

\bibliographystyle{apsrev4-1}


\end{document}